\documentclass[12pt]{article}
\usepackage{latexsym}
\usepackage{epsfig}
\begin{document}

\newcommand{\beq}{\begin{equation}}
\newcommand{\eeq}{\end{equation}}
\newcommand{\bear}{\begin{eqnarray}}
\newcommand{\eear}{\end{eqnarray}}
\newcommand{\half}{{{1}\over{2}}}
\newcommand{\nn}{\nonumber}
\newcommand{\pa}{\partial}

\begin{flushright} KUL-TF-99/18 \\ gr-qc/9905084
\end{flushright}
\vskip 2.5cm
\begin{center}
{\Large \bf A `warp drive' with more reasonable total energy 
requirements} \\
\vskip 1.5cm
{\bf Chris Van Den Broeck$^\dagger$} \\
{\small Instituut voor Theoretische Fysica, \\
Katholieke Universiteit Leuven, B-3001 Leuven, Belgium }
\end{center}
\vskip 4cm
\begin{center}
{\bf Abstract}
\begin{quote}
I show how a minor modification of the Alcubierre 
geometry can dramatically 
improve the total energy requirements for a `warp bubble' that can be used to 
transport macroscopic objects. A spacetime is presented for which the total 
negative mass needed is of the order of a few solar masses, accompanied by a 
comparable amount of positive energy. This puts the warp drive in the
mass scale of large traversable wormholes. 
The new geometry satisfies the 
quantum inequality concerning WEC violations and has the same advantages 
as the original Alcubierre spacetime.
\end{quote}
\end{center}
\vfill
\hrule width 5cm
\vskip 2mm
{\small 
$^\dagger$ chris.vandenbroeck@fys.kuleuven.ac.be}

\section{Introduction}

In recent years, ways of effective superluminal travel (EST) within general 
relativity
have generated a lot of attention \cite{Alcubierre,Krasnikov,
FordPfenning,Olum,EverettRoman}. In the simplest definition of superluminal
travel, one has a spacetime with a Lorentzian metric that is Minkowskian
except for a localized region $S$. When using coordinates such that the
metric is $\mbox{diag}(-1,1,1,1)$ in the Minkowskian region, there should
be two points $(t_1,x_1,y,z)$ and $(t_2,x_2,y,z)$ located outside $S$, such
that $x_2-x_1>t_2-t_1$, and a causal path connecting the two. This was a
definition given in \cite{Olum}. An example is the Alcubierre spacetime
\cite{Alcubierre} if the warp bubble exists only for a finite time. Note that
the definition does not restrict the energy--momentum tensor in $S$. Such
spacetimes will violate at least one of the energy conditions (the weak
energy condition or WEC). In the case of the Alcubierre spacetime, the 
situation is even worse: part of the energy in region $S$ is moving 
tachyonically \cite{Krasnikov,Coule}. The `Krasnikov tube' \cite{Krasnikov}
was an attempt to improve on the Alcubierre geometry. In this paper,
we will stick to the Alcubierre spacetime such as it is. It is not 
unimaginable that some modification of the geometry will make the problem
of tachyonically moving energy go away without changing the other essential 
features, but we leave that for future work. Here we will concentrate 
on another problem.

Alcubierres idea was to start with flat spacetime, 
choose an arbitrary curve, and then deform spacetime in the immediate vicinity
in such a way that the curve becomes a timelike geodesic, at the same time 
keeping most of spacetime Minkowskian. A point on the geodesic is surrounded by
a `bubble' in space. In the front of the bubble spacetime contracts,
in the back it expands, so that whatever is inside is `surfing' through 
space with a velocity $v_s$ with respect to an observer in the Minkowskian
region. The metric is
\beq
ds^2 = - dt^2 + (dx - v_s(t) f(r_s) dt)^2 + dy^2 + dz^2
\eeq
for a warp drive moving in the $x$ direction. $f(r_s)$ is a function which 
for small enough $r_s$ is approximately equal to one, becoming exactly 
one in $r_s=0$ (this is the `inside' of the bubble), and goes to zero for 
large $r_s$ (`outside'). $r_s$ is given by
\beq
r_s(t,x,y,z) = \sqrt{(x-x_s(t))^2 + y^2 + z^2},
\eeq
where $x_s(t)$ is the $x$ coordinate of the central geodesic, which is 
parametrized by coordinate time $t$, and $v_s(t)={{dx_s}\over{dt}}(t)$. A 
test particle in the center of the bubble is not only weightless and travels 
at arbitrarily large velocity with respect to an observer in the large $r_s$ 
region, it also does not experience any time dilatation.

Unfortunately, this geometry violates the strong, dominant, and especially
the weak energy condition. This is not a problem per se, since 
situations are known in which the WEC is violated quantum mechanically, such
as the Casimir effect. However, Ford and Roman \cite{q1,q2,q3,q4} suggested 
an uncertainty--type principle which places a bound on the extent
to which the WEC is violated by quantum fluctuations of scalar and 
electromagnetic fields: The larger the violation, the shorter the time it 
can last 
for an inertial observer crossing the negative energy region. This so--called 
quantum inequality (QI) can be used as a test for the 
viability of would--be spacetimes allowing superluminal travel. 
By making use of the QI, Ford and Pfenning \cite{FordPfenning} were able to
show that a warp drive with a macroscopically large bubble must contain an 
unphysically large amount of negative energy. This is because the QI 
restricts the bubble wall to be very thin, and for a macroscopic bubble the 
energy is roughly proportional to $R^2/\Delta$, where 
$R$ is a measure for the bubble radius and $\Delta$ for its wall thickness. 
It was shown that a bubble with a radius of 100 meters would require a 
total negative energy of at least
\beq
E \simeq - 6.2 \times 10^{62} v_s \,\, \mbox{kg}, 
\eeq
which, for $v_s \simeq 1$, is ten orders of magnitude bigger than the total 
positive mass of the entire visible Universe. However, the same authors also 
indicated that warp bubbles are still conceivable if they are microscopically 
small. We shall exploit this in the following section.

The aim of this paper is to show that a trivial modification of the
Alcubierre geometry can have dramatic consequences for the total negative 
energy as calculated in \cite{FordPfenning}. In section 2, I will explain
the change in general terms. In section 3, I shall pick a specific example
and calculate the total negative energy involved. In the last 
section, some drawbacks of the new geometry are discussed. 

Throughout this note, we will use units such that $c=G=\hbar=1$, except when
stated otherwise.

\section{A modification of the Alcubierre geometry}

We will solve the problem of the large negative energy 
by keeping the {\it surface area} of the warp bubble itself microscopically 
small, while at the same time expanding the spatial {\it volume} inside the 
bubble. The most natural way to do this is the following:
\beq
ds^2 = - dt^2 + B^2(r_s) [(dx - v_s(t) f(r_s) dt)^2 + dy^2 + dz^2].
\label{metric}
\eeq
For simplicity, the velocity $v_s$ will be taken constant. $B(r_s)$ is a 
twice differentiable function such that, for some $\tilde{R}$ and 
$\tilde{\Delta}$,
\bear
B(r_s)=1+\alpha & \mbox{for} & r_s <\tilde{R}, \nn\\
1<B(r_s) \leq 1+\alpha & \mbox{for} & \tilde{R} \leq r_s < \tilde{R} 
+ \tilde{\Delta}, \nn\\
B(r_s)=1 & \mbox{for} & \tilde{R} + \tilde{\Delta} \leq r_s,
\eear
where $\alpha$ will in general be a very large constant; $1+\alpha$ is the
factor by which space is expanded. For $f$ we will choose a function
with the properties 
\bear
f(r_s)=1 & \mbox{for} & r_s < R, \nn\\
0 < f(r_s) \leq 1 & \mbox{for} & R \leq r_s < R + \Delta, \nn\\
f(r_s)=0 & \mbox{for} & R + \Delta \leq r_s, \nn
\eear
where $R > \tilde{R} + \tilde{\Delta}$.
See figure 1 for a drawing of the regions where $f$ and $B$ vary.
\begin{figure}
\begin{center}
\setlength{\unitlength}{1cm}
\begin{picture}(10,7)
\put(0,0){\epsfig{file=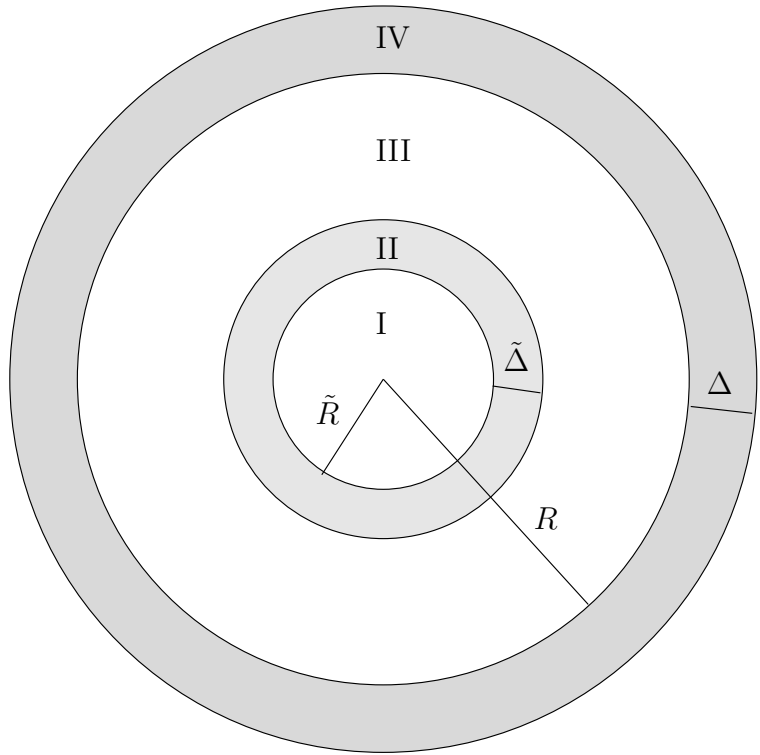,width=10cm}}
\put(4.1,4.4){$\tilde{R}$}
\put(7,3){$R$}
\put(6.6,5.1){$\tilde{\Delta}$}
\put(9.3,4.8){$\Delta$}
\put(4.9,5.6){I}
\put(4.9,6.6){II}
\put(4.9,7.9){III}
\put(4.9,9.4){IV}
\end{picture}
\caption{Region I is the `pocket', which has a large
inner metric diameter. II is the transition region from the blown--up 
part of space to the `normal' part. It is the region where $B$ varies. 
From region III outward we have the original Alcubierre metric.
Region IV is the wall of the warp bubble; this
is the region where $f$ varies. 
Spacetime is flat, except in the shaded regions.}
\end{center}
\end{figure}

Notice that this metric can still be written in the $3+1$ formalism,
where the shift vector has components $N^i = (-v_s f(r_s),0,0)$,
while the lapse function is identically $1$.

A spatial slice of the geometry one gets in this way can be easily visualized
in the `rubber membrane' picture. A small Alcubierre bubble surrounds a neck 
leading to a `pocket' with a large internal volume, with a flat region in the 
middle. It is easily calculated that the center $r_s=0$ of the pocket
will move on a timelike geodesic with proper time $t$.

\section{Building a warp drive}

In using the metric (\ref{metric}), we will build a warp drive with the
restriction in mind that all features should have a 
length larger than the Planck length $L_P$. One structure at 
least, the warp bubble wall, cannot be made thicker than approximately one 
hundred Planck lengths for velocities $v_s$ in the order of 1, as proven in 
\cite{FordPfenning}:
\beq
\Delta \leq 10^2 \, v_s \, L_P. \label{wall}
\eeq
 
We will choose the following numbers for $\alpha$, $\tilde{\Delta}$, 
$\tilde{R}$, and $R$:
\bear
\alpha &=& 10^{17}, \nn\\
\tilde{\Delta} &=& 10^{-15}\, \mbox{m}, \nn\\
\tilde{R} &=& 10^{-15}\, \mbox{m}, \nn\\
R &=& 3 \times 10^{-15} \, \mbox{m}.
\label{numbers} 
\eear
The outermost surface of the warp bubble will have an area corresponding to 
a radius of approximately $3 \times 10^{-15}\, \mbox{m}$, while the inner 
diameter of the `pocket' is 200 m. For the moment, these numbers will seem
arbitrary; the reason for this choice will become clear later on.

Ford and Pfenning \cite{FordPfenning} already calculated the minimum 
amount of negative energy associated with the warp bubble:
\beq
E_{IV} = -{{1}\over{12}} v_s^2 \left( {{(R+{{\Delta}\over{2}})^2}
      \over{\Delta}} + 
      {{\Delta}\over{12}} \right), \label{ef}
\eeq
which in our case is the energy in region IV. The expression is the same 
(apart from a change due to 
our different conventions) because $B=1$ in this region, and the
metric is identical to the original Alcubierre metric. For an 
$R$ as in (\ref{numbers}) and taking 
(\ref{wall}) into account, we get approximately
\beq
E_{IV} \simeq - 6.3 \times 10^{29} v_s \,\, \mbox{kg}.
\eeq

Now we calculate the energy in region II of the figure. In this region, we
can choose an orthonormal frame
\bear
e_{\hat{0}} &=& \pa_t + v_s \pa_x, \nn\\
e_{\hat{i}} &=& {{1}\over{B}}\pa_i \label{frame}
\eear
($i=x,y,z$). 
In this frame, there are geodesics with velocity $u^{\hat{\mu}} = (1,0,0,0)$, 
called 
`Eulerian observers' \cite{Alcubierre}. We let the energy be measured by a 
collection of 
these  observers who are temporarily swept along with the warp drive. Let us
consider the energy density they measure locally in the region II, 
at time $t=0$, when $r_s=r=(x^2+y^2+z^2)^{1/2}$. It is given by
\beq
T_{\hat{\mu}\hat{\nu}} u^{\hat{\mu}} u^{\hat{\nu}} = T^{\hat{0} \hat{0}} 
= {{1}\over{8 \pi}} \left(
         {{1}\over{B^4}}(\pa_r B)^2 
         - {{2}\over{B^3}} \pa_r \pa_r B
         - {{4}\over{B^3}} \pa_r B {{1}\over{r}} \right).\label{density}
\eeq
We will have to make a choice for the $B$ function. It turns out that
the most obvious choices, such as a sine function or a low--order 
polynomial, lead to pathological geometries, in the sense that they have
curvature radii which are much smaller than the Planck length. This is
due to the second derivative term, which is also present in the 
expressions for the Riemann tensor components and which for these functions 
takes enormous absolute values in a very small region near $r=\tilde{R}+
\tilde{\Delta}$. To avoid this, we will choose for $B$ a polynomial
which has a vanishing second derivative at $r=\tilde{R}+\tilde{\Delta}$. 
In addition, we will demand that a large number of derivatives vanish at 
this point. A choice that meets our requirements is
\beq
B = \alpha(-(n-1) w^n + n w^{n-1}) + 1, \label{B}
\eeq
with 
\beq
w = {{\tilde{R}+\tilde{\Delta}-r}\over{\tilde{\Delta}}}
\eeq
and $n$ sufficiently large. 

As an example, let us choose $n=80$. Then one can check that $T^{\hat{0} 
\hat{0}}$ will be negative for $0 \leq w \leq 0.981$ 
and positive for $w>0.981$. It has a strong negative peak at $w=0.349$, where 
it
reaches the value
\beq
T^{\hat{0} \hat{0}} = -4.9 \times 10^2 {{1}\over{\tilde{\Delta}^2}}.
\eeq

We will use the same definition of total energy as in \cite{FordPfenning}: we
integrate over the densities measured by the Eulerian observers as they
cross the spatial hypersurface determined by $t=0$. If we restrict the
integral to the part of region II where the energy density is negative,
we get
\bear
E_{II,-} &=& \int_{II,-} d^3 x \sqrt{|g_S|} T_{\hat{\mu}\hat{\nu}} 
u^{\hat{\mu}} u^{\hat{\nu}} \nn\\ 
&=& 4 \pi \tilde{\Delta} \int_{0}^{0.981} dw (2-w)^2 B(w)^3 
\tilde{T}^{\hat{0}\hat{0}}(w) 
\nn\\
&=& -1.4 \times 10^{30} \mbox{kg}
\eear 
where $\tilde{T}^{\hat{0}\hat{0}}$ is the energy density with length 
expressed in units
of $\tilde{\Delta}$, and $g_S=B^6$ is the determinant of the spatial metric 
on the
surface $t=0$. In the last line we have reinstated the factor $c^2/G$ to
get the right answer in units of kg.
The amount of positive energy in the region $w>0.981$ is
\beq
E_{II,+} = 4.9 \times 10^{30} \mbox{kg}.
\eeq 
Both $E_{II,-}$ and $E_{II,+}$ are in the order of a few solar masses.
Note that as long as $\alpha$ is large , these energies do not vary much
with $\alpha$ if $\tilde{R}=\tilde{\Delta}$ and $\alpha \tilde{R}=100 \, 
\mbox{m}$. The value of $R$ in (\ref{numbers}) is roughly the largest that 
keeps
$|E_{IV}|$ below a solar mass for $v_s \simeq 1$.

We will check whether the QI derived by Ford and
Roman is satisfied for the Eulerian observers. The QI was originally
derived for flat spacetime \cite{q1,q2,q3,q4}, where for massless scalar
fields it states that
\beq
{{\tau_0}\over{\pi}} \int_{-\infty}^{+\infty}
d\tau {{\langle T_{\mu \nu} u^\mu u^\nu \rangle}\over{\tau^2 + \tau_0^2}}
\geq
-{{3}\over{32 \pi^2 \tau_0^4}} \label{qi}
\eeq
should be satisfied for all inertial observers and for all `sampling times' 
$\tau_0$. In \cite{q5}, it was argued that the inequality should also
be valid in curved spacetimes, provided that the sampling time is chosen
to be much smaller than the minimum curvature radius, so that the
geometry looks approximately flat over a time $\tau_0$. 

The minimum curvature radius is determined by the largest component
of the Riemann tensor. It is easiest to calculate this tensor after 
performing a local coordinate transformation $x'=x - v_s t$ in region II, so 
that the metric becomes 
\beq
g_{\mu \nu} = \mbox{diag}(-1,B^2,B^2,B^2). 
\eeq
Without loss of 
generality, we can limit ourselves to points on the line $y=z=0$; in 
the coordinate system we are using, the metric is spherically symmetric and 
has no preferred directions. Transformed to the orthonormal frame 
(\ref{frame}), 
the largest component (in absolute value) of the Riemann tensor is
\beq
R_{\hat{1} \hat{2} \hat{1} \hat{2}} = {{1}\over{B^4}}(\pa_r B)^2 
- {{1}\over{B^3}} \pa_r^2 B
- {{1}\over{B^3}} \pa_r B {{1}\over{r}}.
\eeq
The minimal curvature radius can be calculated using the value of
$R_{\hat{1} \hat{2} \hat{1} \hat{2}}$ where its absolute value is largest, 
namely at $w=0.348$. This 
yields
\bear
r_{c,min} &=& {{1}\over{\sqrt{|R_{\hat{1} \hat{2} \hat{1} \hat{2}}|}}
} \nn\\
        &=& {{\tilde{\Delta}}\over{72.5}} \nn\\
        &=& 1.4 \times 10^{-34} \,\mbox{m},
\eear
which is about ten Planck lengths. (Actually, the choice $n=80$ in (\ref{B})
was not entirely arbitrary; it is the value that leads to the largest
minimum curvature radius.) For the sampling time we choose
\beq
\tau_0 = \beta r_{c,min}, 
\eeq
where we will take $\beta=0.1$. Because $T^{\hat{0} \hat{0}}$ doesn't vary 
much over this time, the QI (\ref{qi}) becomes
\beq
T^{\hat{0} \hat{0}} \geq
-{{3}\over{32 \pi^2 \tau_0^4}}. \label{QInumbers}
\eeq
Taking into account the hidden factors $c^2/G$ on the left and $\hbar/c$ on 
the right, the left hand side is about $-6.6 \times 10^{93} \, 
\mbox{kg}/\mbox{m}^3$ at its smallest, while the right hand side is 
approximately 
$-9.2 \times 10^{94}\,\mbox{kg}/\mbox{m}^3$. We conclude 
that the QI is amply satisfied.

Thus, we have proven that the total energy requirements for a warp drive
need not be as stringent as for the original Alcubierre drive.   

\section{Final remarks}
                   
By only slightly modifying the Alcubierre spacetime, we succeeded in 
spectacularly reducing the amount of negative energy that is needed,
while at the same time retaining all the advantages of the original 
geometry.  The spacetime and the simple calculation I presented should be 
considered as a proof of principle concerning the total energy 
required to sustain a warp drive geometry. This doesn't mean that the
proposal is realistic. Apart from the fact that the total energies are
of stellar magnitude, there are the unreasonably large energy 
{\it densities} 
involved, as was equally the case for the original Alcubierre drive. Even if
the quantum inequalities concerning WEC violations are satisfied, there remains
the question of generating enough negative energy. Also,
the geometry still has structure with sizes only a few orders of magnitude 
above the Planck scale; this seems to be generic for spacetimes allowing 
superluminal travel.

However, what was shown is that the energies needed to sustain a warp 
bubble are much smaller than suggested in \cite{FordPfenning}. This means
that a modified warp drive roughly falls in the mass bracket of a large
traversable wormhole \cite{Visser}. However, the warp drive has trivial
topology, which makes it an interesting spacetime to study.

\medskip
\section*{Acknowledgements}

I would like to thank P.--J. De Smet, L.H. Ford and P. Savaria for very 
helpful comments.

\end{document}